\documentclass[aps,pra,twocolumn,groupedaddress,floatfix,showpacs]{revtex4}

\usepackage{color}
\usepackage{units}
\usepackage{graphicx}
\usepackage{bm}

\begin{document}

\title{Demonstrating coherent control in $^{85}$Rb$_2$ using ultrafast laser pulses: a theoretical outline of two experiments}

\author{Hugo E. L. Martay}
\email{h.martay1@physics.ox.ac.uk}
\affiliation{Clarendon Laboratory, Department of Physics, University of Oxford, Oxford, OX1 3PU, United Kingdom}
\author{David J. McCabe}
\affiliation{Clarendon Laboratory, Department of Physics, University of Oxford, Oxford, OX1 3PU, United Kingdom}
\author{Duncan G. England}
\affiliation{Clarendon Laboratory, Department of Physics, University of Oxford, Oxford, OX1 3PU, United Kingdom}
\author{Melissa E. Friedman}
\affiliation{Clarendon Laboratory, Department of Physics, University of Oxford, Oxford, OX1 3PU, United Kingdom}
\author{Jovana Petrovic}
\affiliation{Clarendon Laboratory, Department of Physics, University of Oxford, Oxford, OX1 3PU, United Kingdom}
\author{Ian A. Walmsley}

\affiliation{Clarendon Laboratory, Department of Physics, University of Oxford, Oxford, OX1 3PU, United Kingdom}

\date{\today}
\begin{abstract}
Calculations relating to two experiments that demonstrate coherent control of preformed rubidium-85 molecules in a magneto-optical trap using ultrafast laser pulses are presented. {The two experiments are a step towards the stabilization of ultracold rubidium dimers using ultrafast lasers.} In the first experiment, it is shown that pre-associated molecules in an incoherent mixture of states can be made to oscillate coherently using a single ultrafast pulse. A novel mechanism that can transfer molecular population to more deeply bound vibrational levels is used in the second. Optimal parameters of the control pulse are presented for the application of the mechanism to molecules in a magneto-optical trap. The calculations make use of an experimental determination of the initial state of molecules photoassociated by the trapping lasers in the magneto-optical trap and use shaped pulses consistent with a standard ultrafast laser system. {The experiment's purpose is to demonstrate and evaluate the use of ultrafast shaped pulses to manipulate ultracold rubidium dimers with a view to the eventual stabilization of the molecules}.
\end{abstract}


\maketitle

\section{Introduction}

	The success in cooling atomic gases to quantum degeneracy and the resulting rich phenomenology has prompted efforts to cool molecular gases to similarly cold temperatures. 
Two alternative approaches have been demonstrated that create cold molecules. The first approach relies on methods that directly cool preformed molecules, but the lack of a closed cooling cycle precludes the use of traditional atomic laser cooling techniques. As a result, molecules have not been directly cooled to the sub-microkelvin temperatures needed for quantum degeneracy.
By contrast, techniques associating precooled atoms retain the very cold centre-of-mass motion of their parent atoms, but the stabilization of the molecules in the lowest rovibrational levels is an ongoing challenge. To achieve a stable ultracold gas, the cold molecules must be transferred to their vibrational, rotational and electronic ground state. Spontaneous decay must be avoided if the very coldest temperatures are to be achieved since spontaneous emission heats the centre-of-mass motion of the molecules.

{Several recent efforts have been successful in stabilizing ultracold molecules. The majority of these rely on stimulated Raman adiabatic passage (STIRAP) \cite{jaksch_2002,winkler_2007}. Typically, highly vibrationally excited molecules are created using a magnetic field sweep across a Feshbach resonance which associates molecules into a highly vibrationally excited state. STIRAP then transfers these to more deeply bound states using pairs of pulses. The process is very efficient, and although there are constraints on the pairs of levels that population may be transferred between, multiple STIRAP transitions may be made that can populate the vibrational ground state. Notable demonstrations of this method include the stabilization of rubidium dimers into their triplet vibrational ground state with a single STIRAP pulse pair \cite{lang_2008}, the stabilization of KRb dimers into their singlet and triplet ground states \cite{ni_2008}, the transfer of caesium dimers using STIRAP to the seventy third vibrational level \cite{danzl_2008, mark_2009}, then to the vibrational ground state \cite{cs_ground_state}.}

{By varying a magnetic field and using a series of stimulated radio-frequency transitions, it is also possible to manipulate diatomic molecules \cite{lang_2008_feshbach}, but without exorbitantly large magnetic fields, it is difficult to transfer molecules to any vibrational state bound by more than a few GHz. This technique can be used to achieve a precise initial state for either STIRAP or some other control mechanism.} 

{A second route to the vibrational ground state of the caesium dimer has recently been demonstrated \cite{viteau_2008}. In this scheme, a continuous wave laser is used to photoassociate molecules into deeply bound vibrational states. 
Broadband pulses are used to excite the deeply bound caesium dimers. The pulses are shaped spectrally to avoid exciting the vibrational ground state. Over several cycles of pump and decay, population begins to accumulate in the vibrational ground state. The disadvantage of this scheme is that it relies on spontaneous emission, and therefore the centre of mass motion of the molecules is heated. Also, the scheme relies on a fortuitous alignment of excited state potentials in Cs$_2$ for an initial injection of population into deeply bound vibrational states. Such a mechanism has not been demonstrated in Rb$_2$, and without some way to populate the lowest vibrational states, the pump-decay cycling results in the dissociation of the dimer with negligible accumulation in any vibrational state.}

{A complementary technique to STIRAP is control using ultrafast laser pulses. Whereas STIRAP in general avoids spontaneous decay by avoiding populating decay-susceptible states, short pulse techniques transfer population on a timescale shorter than the decay time of the excited molecule. The technique uses laser pulses with durations between tens of femtoseconds to several picoseconds, and benefits from the relative ease with which such pulses can be shaped spectrally. On paper, the achievements of short pulse control on ultracold molecules have been modest. Photoassociation by ultrafast laser pulses has been demonstrated \cite{salzmann_2008, mccabe_2009} and the coherent control of photodissociation \cite{brown_2006, salzmann_2006} of preassociated Rb$_2$ has also been demonstrated. Molecular stabilization using ultrafast lasers has not yet been demonstrated.}

{Short pulse control, unlike STIRAP, can address many transitions at once. This allows for a coherent final state to be chosen freely, allowing, for example, the creation of shaped wavepackets. If short pulses were to be used optimally and shaped with sufficient precision, they could be used with far more versatility than STIRAP. 
By contrast, STIRAP must be tailored to a specific transition, and each population transfer requires two phase-locked lasers. 
An experimental setup that achieves transfer across one transition using STIRAP is not able to address any other transition without modification. Short pulse control, by contrast, could change the final occupation of vibrational states in real time and with a far less restricted range of output states. Applying an adaptive technique such as a genetic algorithm \cite{salzmann_2006} to the pulse shaper could then be used to create a desired superposition of states. Short pulse control, coupled with an adaptive system, could be used to manipulate molecules for which detailed spectroscopic information is not present.}
{For these reasons, it may be argued that ultrafast shaped pulses will have an important role in future developments in this field.}  

	 A scheme has been proposed \cite{koch_2006, jordi_2007} that stabilizes loosely bound $^{85}$Rb$_2$ molecules by transferring them to an attractive excited state where they oscillate. A short time later they are transferred back to the electronic ground state. Optimal control pulses for such a `pump-dump' scheme have been suggested \cite{poschinger_2006}.

Although it may be shown theoretically that a sufficiently tailored control pulse could transfer population to the vibrational ground state with high efficiency,
there is uncertainty about how the calculations relate to experiments where the parameters may be controlled with finite precision. In this paper the gap is narrowed by the suggestion of two experiments and realistic calculations relating to them, that demonstrate several of these principles on a readily available source of molecules. {The two experiments represent a step forward in the experimental application of control using ultrafast lasers to ultracold molecules.}

In order for this pump-dump scheme to work, the principles of its operation have to be shown to work: the creation, detection, and manipulation of a suitable initial state; the creation, observation, and control of the oscillating intermediate step; and the measurement of the final state.

Unlike previous work, the calculations here relate to two experiments manipulating preformed molecules whose initial state was determined by analysis of experimental data. 
The experiments are carried out on a dilute gas of $^{85}$Rb atoms trapped in a magneto-optical trap (MOT). The gas cloud contains a reservoir of loosely bound $^{85}$Rb$_2$ molecules due to photoassociation by the MOT trapping lasers and three-body recombination \cite{lozeille_2006, gabbanini_2000, caires_2005}. The experiments manipulate these molecules. The parameters of the control pulses were chosen to correspond to a standard ultrafast laser system.

 The first experiment creates a coherently oscillating excited wavepacket. The experimental demonstration of this would open the door to wavepacket shaping and focussing \cite{koch_2006, koenig_2004, jordi_2007}, that is essential to the pump-dump scheme and other coherent control processes using ultrafast lasers. A short laser pulse is used to excite the ground state (5S + 5S) $^{85}$Rb$_2$ molecules creating a coherently oscillating wavepacket in the excited electronic states associated with the 5S + 5P$_{1/2}$ asymptote. The calculations presented here show that coherent oscillations may be caused in the excited state despite the incoherence of the initial state.

The second experiment demonstrates the population transfer of loosely bound ground state dimers to more deeply bound vibrational levels and is a second way to demonstrate coherent control using ultrafast lasers. 
The process of optimization of the control pulses and the detection of the resulting molecules are similar to the equivalent processes in the pump-dump scheme. The mechanism used to increase the binding energy of the molecules may be applied to other molecular species.

	In Section \ref{theory}, the model used to predict the behaviour of an atom pair under the influence of a time dependent electric field is presented. The initial state of molecules photoassociated by the MOT trapping lasers is given. Sections \ref{exp1} and \ref{exp2} detail the experiments and give predictions of their outcomes and, in the second experiment, the optimal pulse parameters.

\begin{figure}
 \centering
 \includegraphics[width=\columnwidth, angle=0]{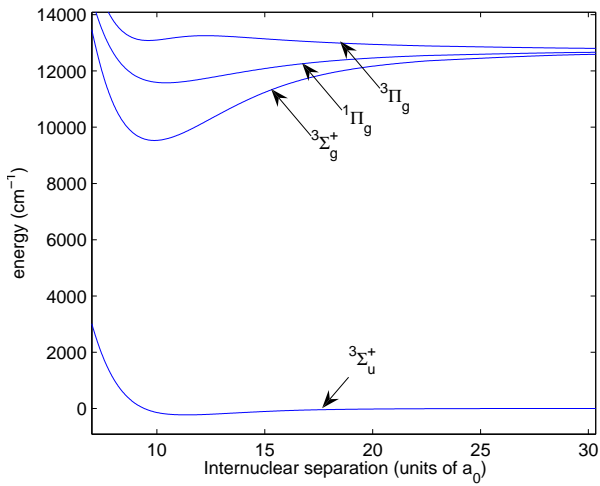}
  \caption{{(Color online) Eight electronic states are used in the majority of the calculations presented for reasons discussed in the Appendix. The eight electronic states in the model used in this paper have four distinct symmetries. Each symmetry has an associated potential energy curve. }The potential energy curves for these four symmetries are shown here. They are taken from references \cite{park_2001,private_jeung,private_greene, klausen_2001}.\label{pecs1}}
\end{figure}

\section{Theory\label{theory}}

\subsection{Model / Excited state dynamics\label{dynamics}}
	A model containing eight electronic states is used to calculate the time dependent wavefunction of the dimer during and after its interaction with a time dependent laser field. The model Hamiltonian takes the form 
\[ H_{j,k} = - \delta_{j,k}{\hbar^2\over 2\mu} {\partial^2\over \partial R^2}  + \delta_{j,k}V_j(R) + \xi(t)D_{j,k} + W_{j,k},\]
	where $j$ and $k$ are the indices of the eight channels. $V_j(R)$ is the potential energy of a given electronic state at a given internuclear separation, $R$. $D_{j,k}$ is the electronic dipole moment of the transition from state $j$ to $k$. $\xi(t)$ is the time varying electric field of the pump pulse.  $W_{j,k}$ is the spin-orbit coupling between electronic states.
	
	The eight electronic states (see Appendix) consist of two ground states with $(a)^3\Sigma_u^+$ symmetry (with zero or one unit of spin angular momentum projected onto the internuclear axis.) and six excited states associated with the 5S + 5P asymptotes, of which three have $^3\Pi_g$ symmetry, two have $^3\Sigma_g^+$ symmetry, and one has $^1\Pi_g$ symmetry. {These eight states form a set that are not coupled by either dipole transitions or spin-orbit coupling to any other states outside the set that are associated with the 5S + 5S or 5S + 5P asymptotes. For reasons discussed in the Appendix, this eight state model was found to give the same results as the full 15 state model.}

	Each symmetry has a distinct potential energy curve. The excited state potential energy curves for the dynamics calculations are taken from \textit{ab initio} calculations \cite{park_2001,private_jeung}. Their long range behaviours are modified to match published dispersion coefficients \cite{bergeman_2006}. The ground state potential energy curves are taken from references \cite{private_greene, klausen_2001}. Spin-orbit couplings and dipole transition strengths are assumed to be constant and take their asymptotic values. 
These were inferred from the energy difference between the 5P$_{3/2}$ and 5P$_{1/2}$ states and their lifetimes \cite{nist_data}. The potential energy curves are shown in Fig.~\ref{pecs1}. The model neglects the rotation of the internuclear axis which takes place on longer timescales than the experiments. The angle between the internuclear axis and the polarization of the laser field is fixed in each calculation. 
 
	The initial state for the calculations is a density matrix, diagonal in the model Hamiltonian's eigenbasis. The diagonal elements are described in Section \ref{initial}. 
	The evolution of the density matrix in time was calculated using a Chebyshev propagator on an analytic mapped grid \cite{talezer_1984, grid}.

\begin{figure}
 \centering
 \includegraphics[width=\columnwidth, angle=0]{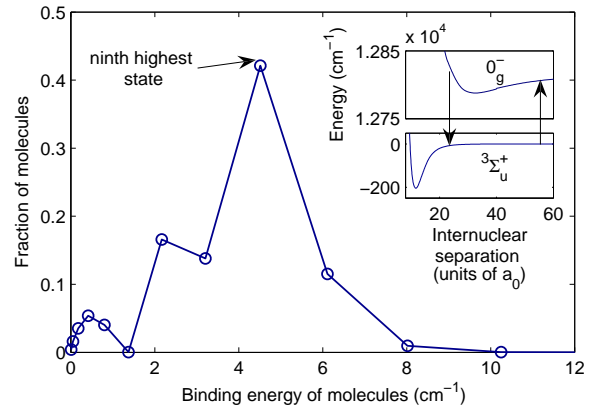}
 \caption{\label{initial_state}(Color online) The relative occupation of triplet bound states in a $^{85}$Rb MOT's molecular reservoir as a function of binding energy. The inset shows the potential energy curves of the $0_g^-$ state (upper) and ground triplet state (lower). The MOT trapping lasers excite atom pairs at long range. Population is concentrated at the inner turning point of the $0_g^-$ potential where it can decay to loosely bound ground triplet molecules with the shown distribution. The relative occupation of each vibrational state was estimated by considering its Frank-Condon overlap with a highly vibrationally excited $0_g^-$ state. {The distribution is insensitive to the choice of vibrational state in $0_g^-$ as long as it is chosen close to dissociation.}
{An RKR potential energy curve \cite{fioretti_2001} was used for the calculation of the $0_g^-$ state.} The distributions agree with the experimentally measured distribution.}
\end{figure}

\subsection{Initial state\label{initial}}
	\noindent Two mechanisms create molecules in a $^{85}$Rb MOT: photoassociation by the trapping lasers, and three-body recombination \cite{gabbanini_2000}. At high trapping laser intensities and lower MOT densities, photoassociation dominates \cite{caires_2005}. In the experimental setup described in this work, the occupation of vibrational levels in the ground triplet electronic state was found to match that predicted assuming photoassociation to be dominant. 

The trapping lasers in a MOT photoassociate unbound atom pairs creating a reservoir of loosely bound ground state molecules. The MOT consists of two lasers tuned to the $D_2$ (5S --- 5P$_{3\over2}$) transition in rubidium. To photoassociate, these must excite atom pairs to electronic states which are both attractive and associated with the 5S + 5P$_{3\over2}$ threshold. Of the states that satisfy both criteria, photoassociation to the $0_g^-$ state is dominant. This mechanism populates the top ten ground triplet vibrational states with a binding energy of up to \unit[4.5]{cm$^{-1}$}. The occupation of each vibrational state in the triplet ground electronic states was estimated by calculating their Frank-Condon overlap with a highly vibrationally excited $0_g^-$ state. 

{Since the trapping lasers responsible for the photoassociation are tuned to within a few MHz of the dissociation threshold, the  $0_g^-$ vibrational state responsible for the photoassociation must be close to the dissociation threshold. It was found that the distribution of population in the ground state vibrational levels resulting from the decay of an $0_g^-$ state was insensitive to the choice of $0_g^-$ state, as long as it was close to the dissociation threshold.}

{An RKR potential energy curve \cite{fioretti_2001} was used for the initial state calculation.}
  The occupation of ground state vibrational states as a function of binding energy is shown in Fig.~\ref{initial_state} with the relevant potential energy curves. The number of molecules in this distribution may be increased by using a continuous wave photoassociation laser tuned to photoassociate unbound atom pairs to the same $0_g^-$ state.

	The occupation of each vibrational state may be measured by molecular spectroscopy experiments using a narrow-band laser tuned to around \unit[14500]{cm$^{-1}$} \cite{lozeille_2006, fioretti_2001}. In an accompanying experimental work to this paper \cite{mccabe_2009}, a pulsed dye laser 
was scanned over a wavelength interval from \unit[14200]{cm$^{-1}$} to \unit[14600]{cm$^{-1}$}. The number of molecular ions produced (via resonantly enhanced multiphoton ionization through the $(2)^3\Sigma_g^+$ state) as a function of wavelength was used to infer the population distribution of the vibrational levels of the molecular ground state. Good agreement was found between experiment and theory, suggesting that the detectable molecules were indeed created by photoassociation to the $0_g^-$ state.

\section{Coherent oscillations\label{exp1}}

	The reservoir of preformed molecules within a $^{85}$Rb MOT are an accessible basis for a coherent control experiment.  However, they are initially in an incoherent mixture of the top ten vibrational states. It is not clear therefore whether or not they can be manipulated coherently. 

\begin{figure}
 \centering
 \includegraphics[width=\columnwidth, angle=0]{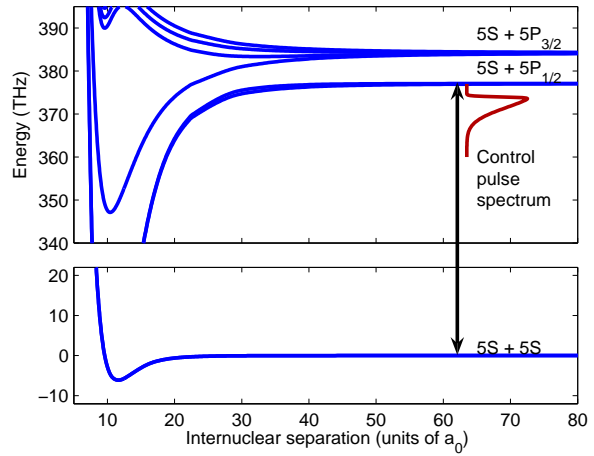}
 \caption{(Color online) The spectrum of the excitation pulse for the oscillation experiment in relation to the adiabatic potential energy curves, showing the two 5S + 5P asymptotes. The pulse is spectrally cut at \unit[374]{THz} to prevent excitation of atoms to the 5P$_{1/2}$ state, and to prevent scattering pairs from being excited. 
\label{diagonal_V}}
\end{figure}

The model described in Section \ref{dynamics} was initiated with the density matrix described in Section \ref{initial}. A time varying electric field was used that approximates an experimentally realizable laser pulse: a Gaussian pulse with a centre frequency of \unit[375]{THz} (\unit[799.2]{nm}), a full width at half maximum intensity of \unit[8]{THz} (\unit[17]{nm}), and spectrally cut to remove intensity at a higher frequency than \unit[374.7]{THz} (\unit[800]{nm}),
removing spectral intensity at the 5S --- 5P$_{1/2}$ transition. The spectral cut prevents atoms from being excited in the MOT which depletes the MOT and floods the ion detectors with atomic ions. Having the spectral cut at \unit[800]{nm} rather than \unit[794.7]{nm} (the transition wavelength) also prevents the photoassociation of unbound atom pairs to long range molecules which oscillate more slowly. The spectrum of the cut pulse as well as the potential energy curves of the excited states are shown in Fig.~\ref{diagonal_V}. Calculations were performed at a range of fluences up to \unit[300]{Jm$^{-2}$} with fluences in both the linear and non-linear regimes.

\begin{figure}
 \centering
 \includegraphics[width=\columnwidth, angle=0]{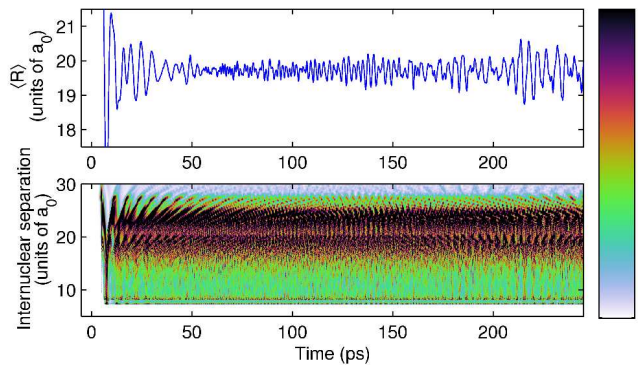}
 \caption{(Color online) The average internuclear separation and population density in the excited electronic states after being excited by a spectrally-cut short Gaussian pump pulse. Higher population density is represented with darker colors. Oscillations with a period of around \unit[5]{ps} can be seen. The oscillations persist with no appreciable decay to at least \unit[250]{ps}. A fractional revival of the vibrational wavepacket is evident at \unit[230]{$a_0$}. 
\label{results1}}
\end{figure}

\subsection{Results}
	For all fluences, population is excited predominantly at an internuclear separation of \unit[20]{$a_0$}. It then oscillates with a period of around \unit[5]{ps}. The oscillations persist to the end of the simulation at \unit[250]{ps}.  
The population density and average internuclear separation with an incident pulse fluence of \unit[20]{Jm$^{-2}$} 
are shown in Fig.~\ref{results1}. Oscillations could be detected if a position sensitive measurement of the excited state population could be made. 

\subsection{Detection of oscillations}
	The ionization cross-section has, in some molecular species, been shown to be dependent on the internuclear separation of the dimer \cite{ionisation_NaI, ionisation_Na2}. It is therefore possible that a suitable ionizing probe pulse can be used to observe the oscillations: The molecular ion signal would vary as a function of the time delay between the pulse causing the oscillation and the probe pulse, revealing details of the oscillations. 

The experiment these calculations imply is similar to a previous experiment \cite{salzmann_2008} in which a magneto-optical trap was illuminated by a cut Gaussian laser pulse, and then probed by a time delayed probe pulse. The molecular ion signal observed consisted of molecules that had been photoassociated by the control pulse rather than preformed molecules. By contrast, the experiment here has a control pulse which is spectrally cut further to the red to extinguish this second source of molecular ions. A continuous wave photoassociation laser tuned to a transition from unbound atoms to bound molecules in the $0_g^-$ electronic state associated with the 5S + 5P$_{3/2}$ asymptote would increase the number of molecules in the initial state used here. This would also help to reduce the relative contribution to the molecular ion signal of molecules photoassociated from unbound atoms by the control pulse.

\section{Population transfer\label{exp2}}

\begin{figure}[t]
 \centering
 \includegraphics[width=\columnwidth, angle=0]{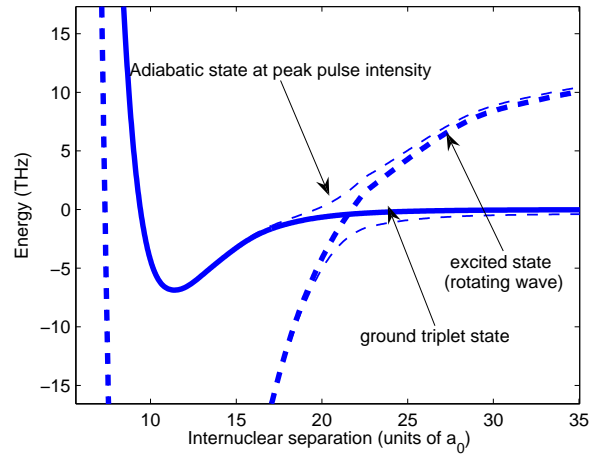}
 \caption{\label{adiabatic_crossing}(Color online) The ground and excited states in the rotating wave approximation, showing how the mechanism described in Section~\ref{mechanism} works. The ground state (solid line) is coupled by a strong electric field to an excited state (thick dashed line). The coupled Hamiltonian can be diagonalized at each internuclear separation, resulting in two adiabatic potential energy curves exhibiting an avoided crossing (thin dashed line). Near the crossing, both states are more attractive than the ground state. Population is not transferred to the excited state. This gives any population near the crossing a momentum kick inwards.}
\end{figure}

\begin{figure}[t]
 \centering
 \includegraphics[width=\columnwidth, angle=0]{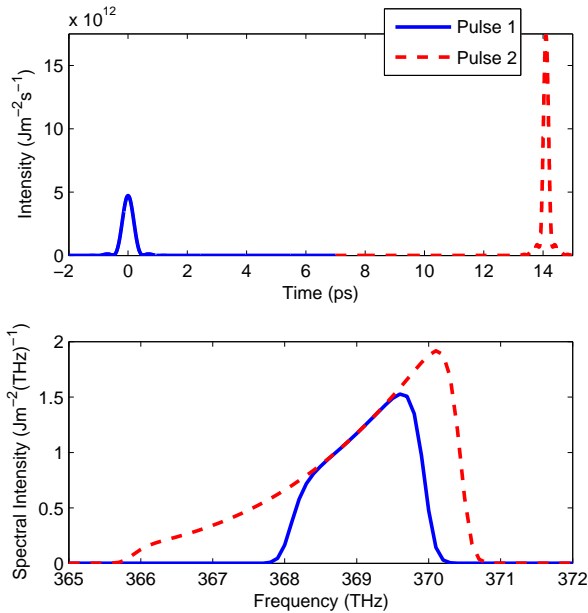}
 \caption{\label{pulseparams}(Color online) The temporal and spectral intensities of a pulse pair that achieves 15\% transfer from the ninth highest to the twelfth highest vibrational level in the triplet ground state. In a similar scheme, two identical pulses may be used with the parameters of the first pulse here. This simpler scheme results in 10\% transfer instead. The parameters of this pulse are given in Table~\ref{opt. parms}}
\end{figure}

\begin{table}  
\begin{tabular}{ l  ll  l}
	\hline
	\hline
	Parameter && Value & \\
	\hline
  	Fluence before shaping && \unit[100]{Jm$^{-2}$} &\\  
	Fluence after shaping && \unit[6.17]{Jm$^{-2}$}&\\  
	Delay && \unit[14.1]{ps} & \\
  	Blue cut (pulse 1) && \unit[369.8]{THz} &(\unit[810.5]{nm})  \\   
	Red cut (pulse 1) && \unit[368.0]{THz} &(\unit[814.4]{nm})\\ 
	Blue cut (pulse 2) && \unit[370.3]{THz} &(\unit[809.4]{nm})  \\  
	Red cut (pulse 2) && \unit[365.8]{THz} &(\unit[819.3]{nm})\\ 
	Chirp 1          && 0&\\
	Chirp 2          && 0&\\
	\hline
\end{tabular}

\caption{\label{opt. parms} Pulse parameters for a pulse pair that achieves around 15\% transfer from the ninth highest vibrational state to the twelfth highest in the ground triplet electronic state.}
\end{table}

	The reservoir of molecules may be expected to have predominantly $^3\Sigma_u^+$ symmetry. Pump-dump schemes as mentioned above are unlikely to work for this symmetry in rubidium due to the lack of any suitable excited potential energy curves.
	Instead, a scheme is proposed here that does not rely on an alignment of excited potential energy curves. The scheme increases the binding energy of the molecules using a dynamic Stark shift. Calculations presented here suggest that the scheme is robust enough with respect to the control field that it is experimentally accessible.

\subsection{Mechanism\label{mechanism}}

	A mechanism is used to increase the binding energy of the dimer. It makes use of a momentum kick as discussed in references \cite{koenig_2007, brown_theory_paper}. The molecules are initially in a stationary wavefunction with an electronic configuration that has very weak attraction between the atoms. An oscillating electric field is applied on a timescale much shorter than the oscillation time of the internuclear separation, but longer than the oscillation time of the valence electrons. This adiabatically mixes the initial electronic configuration with another electronic configuration with stronger attraction between the atoms as shown in Fig.~\ref{adiabatic_crossing}. The population which was in the ground state follows the dressed initial state adiabatically and remains in the ground state after the pulse. Because the dressed state is transiently attractive due to the position-dependent Stark shift, the dimer is given a momentum kick inwards. 

	In the scheme presented here, the molecules then bounce against the inner turning point of the ground state potential. They are given a further momentum kick inwards at a later time by a second control pulse. This slows them down, trapping them in vibrational states with lower energy than their initial state. The technique is applicable wherever two electronic states with different potential gradients are coupled by an electric field. Although the calculations presented here relate to molecules produced in a MOT, the scheme could also be extended to the loosely bound molecules produced by sweeping a magnetic field across a Feshbach resonance.  

	In this application of the technique, the mechanism relies on the interaction between the ground state $^3\Sigma_u^+$ and the excited state with $0_g^-$ symmetry associated with the 5S + 5P$_{1\over2}$ asymptote. This was determined by running the simulations with further reduced models containing different combinations of states (see Appendix).

\subsection{Pulse optimization}

	The transfer of population between an initial state and a more deeply bound target state was optimized. The initial state in the optimization was chosen to be the ninth highest vibrational state --- which is predicted to have the highest occupation as shown in Fig.~\ref{initial_state} --- rather than a density matrix, in order to reduce the computation time of the calculations. The target state was chosen to be the twelfth highest vibrational state. The twelfth highest state was chosen because its binding energy is sufficient to distinguish it from any of the initially populated states, but is similar enough to the initial state that a detection scheme that had been demonstrated to work for the initial state would be likely to work still.

\begin{figure*}
 \centering
 \includegraphics[width=\textwidth, angle=0]{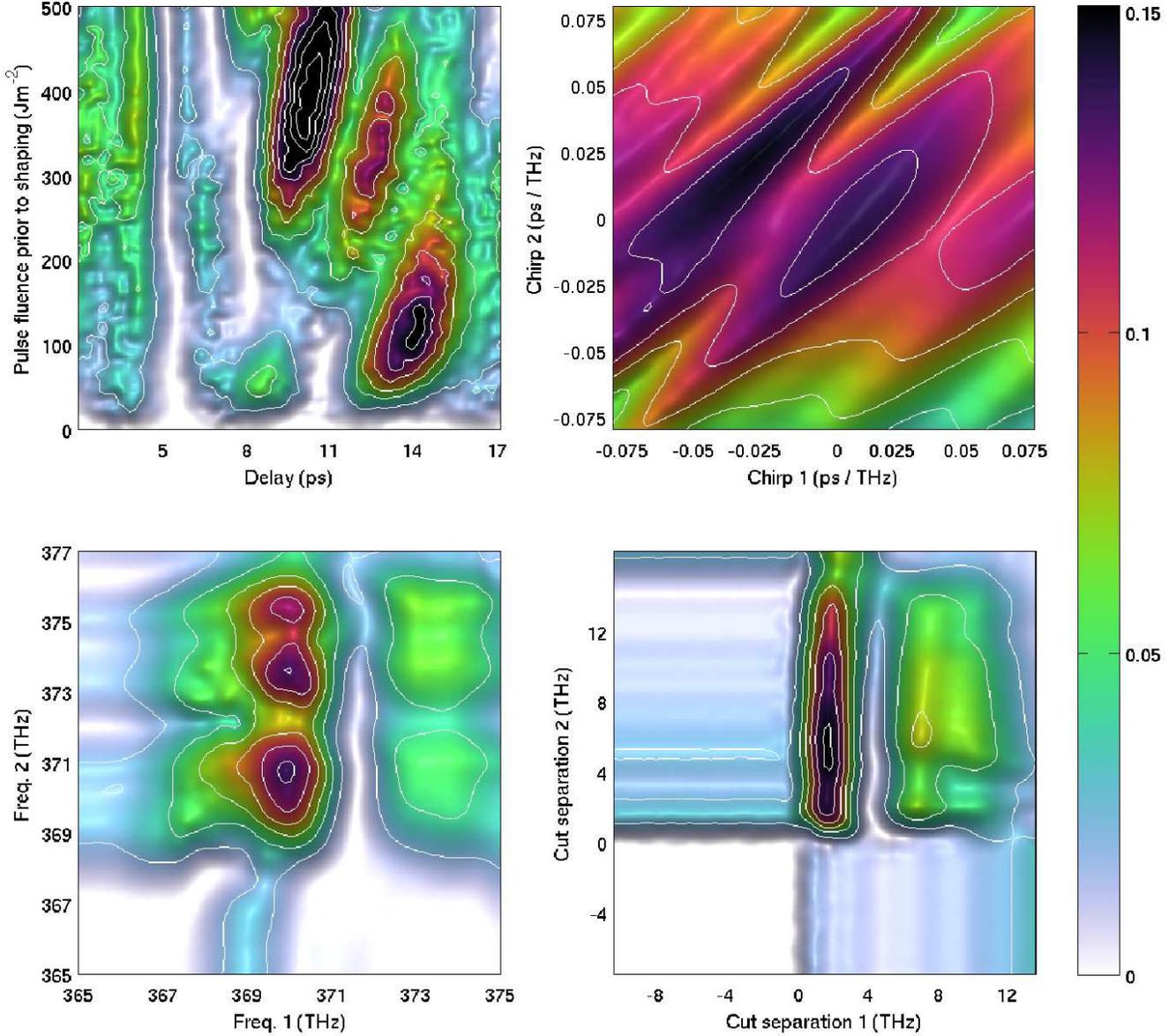}
 \caption{\label{eightD}(Color online) The efficiency of the pulse pair as a function of its eight parameters. The dependence on the fluence of the pulses and the delay between them is shown in the top left graph. The dependence on chirp of the two pulses is shown in the top right. The dependence on the centre frequency of the two pulses and their bandwidths is shown in the bottom two graphs. Negative bandwidth of a pulse results in the pulse being extinguished. Contours represent transfer efficiencies in steps of 2.5\%. Higher efficiencies are represented by darker colors. Here, efficiency is defined by the transfer fraction from the 9th highest vibrational state to the 12th.}
\end{figure*}

	Two laser pulses are used as the control field. The pulses mimic an experimental setup where a Gaussian laser pulse [with a central frequency of \unit[375]{THz} (\unit[799.2]{nm}) and a full width at half maximum intensity of \unit[8]{THz} (\unit[17]{nm})] is split into two equal pulses. Each pulse is separated into its Fourier components and is spectrally cut, removing intensity above a threshold wavelength and below a second threshold. Separate chirps were applied to each pulse. The pulse pair therefore has eight parameters: the delay, the wavelengths of the two blue and two red cuts, the fluence of the pulse pair before the spectral cuts, and the chirp of each pulse. The spectral cut has a sigmoid profile with a constant width in accordance with experimental capabilities.

	A genetic algorithm was employed to find a maximum in the transfer efficiency at a series of different fluences. These were examined and the most suitable --- judged by having relatively high transfer efficiencies, low peak laser fields and low fluences --- was taken as the optimal solution. Low fluences allow the control field to be applied over a larger cross-sectional area of the gas cloud in the magneto-optical trap. High peak intensities are undesirable because they increase the likelihood of multiphoton ionization during the control phase of the experiment, and also introduce Stark shifts that are not taken into account in the models used here.

\subsection{Results}
	
At \unit[100]{Jm$^{-2}$} (unshaped), 15\% population transfer from the ninth to the twelfth highest vibrational state was achieved. The pulse parameters are listed in Table \ref{opt. parms}, and their spectra and temporal profiles are given in Fig.~\ref{pulseparams}.

\begin{figure}
 \centering
 \includegraphics[width=\columnwidth, angle=0]{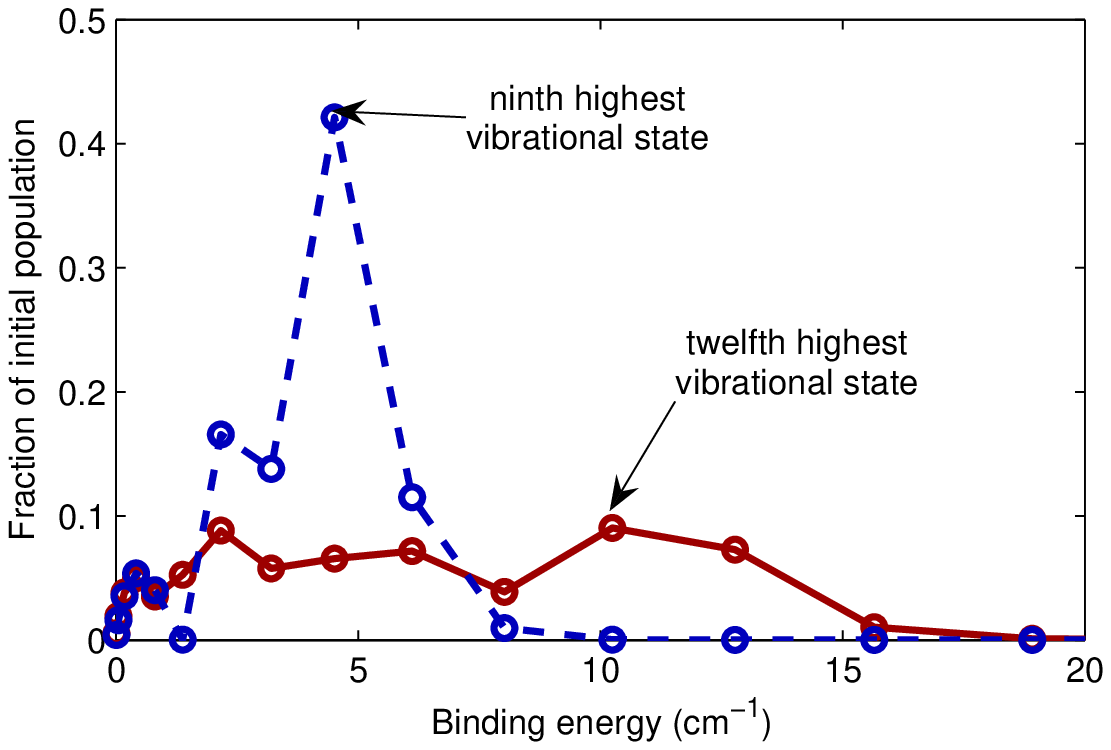}
 \caption{\label{new_distribution}(Color online) The initial (dashed) and final (solid) populations of the highest vibrational states in the ground triplet before and after a control pulse pair. The initial population results from a decay from a high lying $0_g^-$ state, and the final distribution shows how the pulse pair transfers population to more deeply bound states.}
\end{figure}

\begin{figure}
 \centering
 \includegraphics[width=\columnwidth, angle=0]{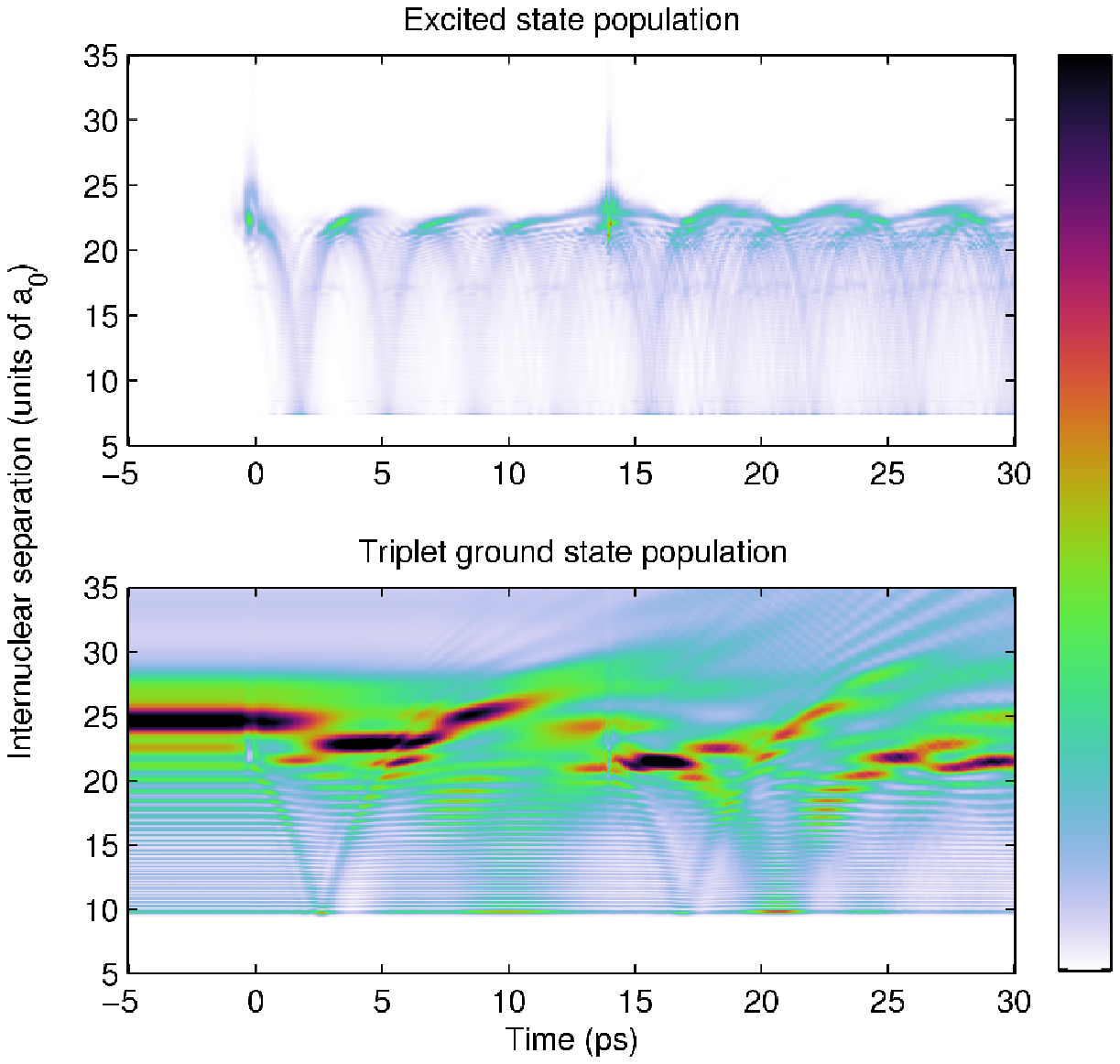}
 \caption{\label{throw_catch_density}(Color online) The population density as a function of time and internuclear separation. The ground (excited) state population is in the lower (upper) graph. Higher population densities are represented by darker colors. Population is in a stationary distribution initially. At \unit[0]{ps}, it is given a momentum kick that results in a bounce against the inner turning point of the triplet potential at roughly \unit[10]{ps}. At \unit[14]{ps} it is stabilized by a second momentum kick. Although the wavefunction in the ground state is strongly affected by the pulses, at no point is a significant population transferred to the excited state. The parameters of the control field are given in Table~\ref{opt. parms}.}
\end{figure}

	The sensitivity of the transfer efficiency to the pulse parameters was established by taking four two-dimensional cross sections through the eight-dimensional search space and plotting the transfer efficiency as a function of each pair of parameters as shown in Fig.~\ref{eightD}. Although the interpretation of the plots is difficult, it can be seen that the scheme is most sensitive to the parameters of the first pulse and the delay between the pulses. The optimal delay corresponds to the time taken for the wavepacket to bounce against the inner turning point of the ground triplet potential and to return to the outer turning point of the target state, as can be seen from the population dynamics in Fig.~\ref{throw_catch_density}.

The efficiency does not change dramatically as a result of nanometer shifts in the centre wavelengths of either pulse or in the bandwidth of the second pulse. The efficiency is sensitive to shifts in the bandwidth of the first pulse. 
It was found that the efficiency decreased roughly linearly with fluence for fluences less than the recommended fluence. 
The scheme is largely insensitive to the details of the second pulse, and it was found that replacing the second pulse with a replica of the first, still with \unit[14.1]{ps} delay, only decreased the efficiency of the scheme by one third.

	The bandwidth of the pulses used in the computation were limited to more than one nanometer by the width and shape of the cutting function. Therefore, it is possible that with a sharper cutting function, narrower bandwidths for the first pulse might be more efficient. 

	The calculations were extended to model how each of the initially populated vibrational levels respond to the control field. An initial density matrix as described in section~\ref{initial} was propagated to obtain a density matrix describing the molecular reservoir after the control pulses. As for the pure initial state, a significant fraction of the population was transferred to vibrational states more deeply bound than the initial distribution. The occupation of vibrational states before and after the pulse can be seen in Fig.~\ref{new_distribution}. The population density for the dimers as a function of time and internuclear separation is shown in Fig.~\ref{throw_catch_density}.

\subsection{Experimental considerations}
	This calculation implies an experiment in which a $^{85}$Rb MOT is illuminated by a shaped pulse pair and then probed via resonantly enhanced multiphoton ionization using a tuneable pulsed dye laser. A demonstrated detection scheme \cite{mccabe_2009} involves scanning the dye laser from \unit[14420]{cm$^{-1}$} to \unit[14480]{cm$^{-1}$} to ionize the triplet ground state molecules. The number of molecular ions produced as a function of probe wavelength measures the occupation of vibrational levels in the triplet ground state.

	The experiment depends on being able to make quite precisely shaped pulses and the fluences required are high, although experimentally possible. It is not clear how experimental considerations such as the presence of third order dispersion and other optical artifacts will affect the transfer efficiency.  
	In turn, the calculations presented here do not take into account losses to states associated with the 5P + 5P asymptotes or multiphoton ionization which could reduce the number of molecules that are transferred to more deeply bound states.  These losses are endemic to experiments using ultrafast laser pulses on molecules. Stark shifts caused by coupling to higher lying electronic states are also neglected.

\section{Conclusion}
	Calculations relating to two experiments are presented here. The calculations show that the first experiment causes coherent oscillation in an incoherent mixture of pre-associated rubidium molecules with a typical period of \unit[5]{ps}.

	A generalizable and experimentally realizable mechanism was found that increases the binding energy of the population of pre-associated molecules. A second experiment was proposed that uses this mechanism with optimized pulse parameters within experimentally realizable limits. 
 	
	The two experiments would test and demonstrate specific techniques such as shaped control pulses and time resolved detection of molecular oscillation in ultracold molecules, processes that will play a vital role in pump-dump schemes to stabilize ultracold molecules. More generally, these experiments are conservative ways to test the fundamental principle that shaped ultrafast laser pulses can be used to coherently manipulate ultracold molecules.

\begin{acknowledgments}
	H. E. L. M  would like to thank Dr.~Thorsten K\"{o}hler for supervision and advice. The authors are grateful to Jordi Mur-Petit for valuable advice and insightful discussion. Financial support is acknowledged from the EPSRC, grant number EP/D002842/1.
\end{acknowledgments}

\section*{APPENDIX}
{The rubidium dimer is modelled in this work by projecting the electronic state onto a finite number of basis states but allowing the nuclear motion to evolve freely. The electronic basis states may be chosen to be eigenstates of the electron Hamiltonian with or without the spin-orbit contributions. The intricacies of Hund's cases, which specify the electronic basis set, are beyond the scope of this manuscript. The reader is referred to Ref.~\cite{herzberg}.}

The model used in the calculations presented here contains eight electronic states. A fuller model containing fifteen states and reduced models were also used for comparison. 

Rb$_2$ has four ground electronic states. One has singlet character and three have triplet character.
Since the specific detection scheme used in this paper detects only triplet population, the calculations exclude the singlet ground state and the excited states not coupled to the ground triplet state.
 Of the triplet ground states ($^3\Sigma_u^+$), one has $0_u^-$ symmetry, one has $1_u^+$, and one has $1_u^-$ symmetry in Hund's case (c) notation, giving the symmetry under a reflection in a plane containing the two nuclei and the control field polarization axis as the superscripted symbol. All three triplet states may be populated by the $0_g^-$ decay mechanism. 

An electric field couples the triplet ground states to twelve excited states, {given in Table~\ref{state_table}. A} selection rule for the interaction with a linearly polarized control field, {and also for spin-orbit coupling} is that the reflection symmetry {of the molecule} does not change, and so molecules may be divided into those with negative and those with positive reflection symmetry.

\begin{table}  
\vskip 0.2cm
\begin{tabular}{c|c|c}
	\hline
	\hline
	Hund's case (c) & Hund's case (a)& Inclusion\\
			& components& in 8 state model\\	\hline
	&ground states&\\
	$0_g^+$ & $^1\Sigma_g^+$ & no\\  
	$0_u^-$ & $^3\Sigma_u^+$ & yes\\
	$1_u^+$ & $^3\Sigma_u^+$ & no\\
	$1_u^-$ & $^3\Sigma_u^+$ & yes\\
	\hline
	&gerade excited states&\\
	$0_g^+$ & $^1\Sigma_g^+$, $^3\Pi_g$ & no\\  
	$0_g^-$ & $^3\Sigma_g^+$, $^3\Pi_g$ & yes\\
	$1_g^+$ & $^3\Sigma_g^+$, $^1\Pi_g$, $^3\Pi_g$ & no\\
	$1_g^-$ & $^3\Sigma_g^+$, $^1\Pi_g$, $^3\Pi_g$ & yes\\
	$2_g^+$ & $^3\Pi_g$ & no\\
	$2_g^-$ & $^3\Pi_g$ & yes\\
	\hline
	\hline
\end{tabular}
\caption{\label{state_table}{There are four ground electronic states of Rb$_2$, and twelve excited gerade electronic states. The triplet states are coupled by an electric field to gerade excited states. If the field is linearly polarized, and the plane of reflection that defines the basis set is chosen to contain this field, then a selection rule exists that the reflection symmetry does not change. For reasons detailed in the text, the singlet ground state is excluded from the model as are all the states with positive reflection symmetry, giving the model eight electronic states. A fuller model containing all fifteen states (not the singlet ground state) was used for comparison.}}
\end{table}

 The dynamics in the excited states with $1_g^+$ and $1_g^-$ symmetry are identical as are the $2_g^+$ dynamics to the $2_g^-$ dynamics whereas the $0_g^-$ dynamics are not the same as the $0_g^+$ dynamics. It was found empirically that omitting the states with positive symmetry did not change the excited state dynamics significantly, because the set of excited states with positive reflection symmetry do not contain any excited population on a potential energy curve that is not also found in the set of states with negative reflection symmetry ---  both states with $0_g^+$ symmetry are repulsive and so are only populated by laser intensity blue of the atomic 5S --- 5P$_{1/2}$ transition which in both experiments was filtered out. It was also found that the mechanism used in the second experiment relies on the $0_g^-$ states and so only works for molecules with negative reflection symmetry. The states with positive reflection symmetry can therefore be omitted from the model because for both experiments the reduced model (8 states) gives the same dynamics as the full model (15 states).

Section~\ref{mechanism} discusses calculations making use of further reduced models. These contain the ground triplet states and the excited states with only one Hund's case (c) symmetry each. Whether or not each minimal model recovers the behaviour of the 8 state model determines which Hund's case (c) states are important in a given calculation.

\end{document}